\documentclass[fleqn,usenatbib]{mnras}

\usepackage{newtxtext,newtxmath}

\usepackage{caption}
\usepackage{subcaption}

\usepackage[T1]{fontenc}

\DeclareRobustCommand{\VAN}[3]{#2}
\let\VANthebibliography\thebibliography
\def\thebibliography{\DeclareRobustCommand{\VAN}[3]{##3}\VANthebibliography}

\usepackage{soul}

\usepackage{graphicx}	
\usepackage{amsmath}	
 
\usepackage{amssymb}	

\usepackage{booktabs}
\usepackage{threeparttable}

\usepackage{color}
\usepackage{ulem}

\title[Polarization for 4 mode-changing pulsars]{Distinct polarization properties for two emission states of four pulsars}

\author[Yan et al.]{
Yi Yan,$^{1,2}$ 
 P.~F. Wang,$^{1,2}$\thanks{E-mail: pfwang@nao.cas.cn}
 and J.~L. Han$^{1,2,3}$\thanks{E-mail: hjl@bao.ac.cn}
\\
$^{1}$National Astronomical Observatories, Chinese Academy of Sciences, Jia20 Datun Road, Beijing 100012, China\\
$^{2}$Astronomy School, the University of Chinese Academy of Sciences,  Beijing 100049, China\\
$^{3}$The FAST key laboratory, Chinese Academy of Sciences, 
Beijing 100012, China
}

\date{Accepted XXX. Received YYY; in original form ZZZ}

\pubyear{2023}

\begin{document}
\label{firstpage}
\pagerange{\pageref{firstpage}--\pageref{lastpage}}
\maketitle

\begin{abstract}
Four pulsars, PSRs J1838+1523, J1901+0510, J1909+0007 and J1929+1844, are found to exhibit bright and weak emission states from sensitive FAST observations. New FAST observations have measured their polarization properties for the two states, and revealed that the polarization profiles, linear polarization percentage, and polarization position angle curves, as well as circular polarization percentage are partially or entirely different in the two emission states. 
Remarkably, PSR J1838+1523 has very different slopes for the polarization position angle curves. PSR J1901+0510 has a wider profile and a higher linear polarization in the weak state than those in the bright state. PSR J1909+0007 has very distinct polarization angle curves for the two modes. While
in the case of PSR J1929+1844, the central profile component evolves with frequency in the bright state, and the senses of circular polarization are opposite in the two modes. The different polarization properties of the two emission states provide valuable insights into the physical processes and emission conditions in pulsar magnetosphere. 
\end{abstract}

\begin{keywords}
pulsars -- individual (J1838+1523, J1901+0510, J1909+0007 and J1929+1844)  
\end{keywords}


\section{Introduction}           
\label{sect:intro}

The mean pulse profile of a pulsar is obtained by averaging hundreds of individual pulses, which is usually stable and represents an unique feature of the pulsar. Soon after the pulsar discovery, the pulse profile of PSR B1237+25 is found not always to have the same profile, and it sometimes switches to an ``abnormal shape'' \citep{Backer1970_ModeChange}. The switching between two or more types of profiles was termed as ``mode-changing'', indicating that the emission modes change for some durations. Up to now, the mode-changing phenomenon has been observed for dozens of pulsars \citep[e.g. PSRs B0031-07,B1237+25,J2321+6024,][]{Huguenin1970,Wang2022,Rahaman2021}. 
\begin{table*}
\centering
\caption[]{Parameters of the FAST pulsar observations. Column (1): pulsar name; Column (2)-(3): pulsar period and dispersion measurement (DM); Column (4): date of observation in the format of yyyymmdd; Column (5): cover name and the beam number of FAST pulsar observations. The mark * in this column indicates the FAST released achieved data for open usage; Column (6): pulsar coordination offset from the beam center in arcminute; Column (7): length of observation time in minutes; Column (8): references for mode-changing observations: [0]: this paper; [1]:\citet{Ferguson1981}; [2]:\citet{Nowakowski1994}; [3]:\citet{Weisberg1986} }
\begin{tabular}{lcrclccc}
  \hline\noalign{\smallskip}
Name         & P0      & DM    & ObsDate  & Cover\_Beam   & offset     & T$_{\rm obs}$ & Ref. \\
                     & (s)     & (cm$^{-3}$pc) &  &                       & ($'$)   & (min)         &     \\
(1)                  & (2)     & (3)   & (4)      & (5)            & (6)    & (7)     & (8)           \\
\hline\noalign{\smallskip}     
J1838+1523           & 0.549 & 67.9  & 20210808 & G45.26+9.83\_M09P3  & 0.80    & 5   &   0    \\
J1901+0510           & 0.615 & 434.1 & 20210502 & J1901+0510\_M01P1*  & 0.01    & 27  &    0   \\ 
                     &       &       & 20210903 & J190129+051533\_M09P1  & 1.67    & 15  &     \\
J1909+0007 (B1907+00)  & 1.017 & 113.0 & 20210701 & G35.14-3.81\_M11P1  & 0.87    & 5   &  0     \\
J1929+1844 (B1926+18)  & 1.220 & 112.5 & 20210111 & J1928+1839\_M19P1  & 0.42    & 30  & 1,2,3 \\
                     &       &       & 20220602 & J192838+183935\_M19P1  & 0.42    & 15  &       \\
\noalign{\smallskip}\hline
\end{tabular}
\label{table:obs}
\end{table*}

\begin{table*}
\setlength\tabcolsep{5.5pt}
\centering
\caption[]{The fractional linear and circular polarization of two emission states for four pulsars. Column (1): pulsar name; Column (2): observation date; Column (3)-(5): the fractional linear, fractional absolute circular and fractional circular polarization L/I, $|V|$/I, V/I of the whole integrated pulse profiles; Column (6)-(8): L/I, $|V|$/I and V/I of the weak emission state; Column (9)-(11): L/I, $|V|$/I and V/I of the bright state. For J1909+0007, `weak' or `bright' are distinguished by the central component.}
\label{table:PolFrac}
 \begin{tabular}{ll|ccc|ccc|ccc}
  \hline\noalign{\smallskip}
JName  &  ObsDate  & \multicolumn{3}{c}{The whole profile} & \multicolumn{3}{c}{Weak state} & \multicolumn{3}{c}{Bright state} \\
\hline\noalign{\smallskip}
	   &     &  $L/I$ (\%)  & $|V|/I$ (\%) & $V/I$ (\%)  & $L/I$ (\%) & $|V|/I$ (\%) & $V/I$ (\%)  & $L/I$ (\%)  & $|V|/I$ (\%) & $V/I$ (\%) \\
(1)    & (2)       & (3)   & (4)     & (5)  & (6) & (7)     & (8)   & (9) & (10)    & (11) \\    
\hline\noalign{\smallskip}
J1838+1523  & 20210808 & 24.6(3) & 19.2(4) & -16.8(4)  & 27.9(5)  & 22.0(8)  & -20.2(8)  & 21.1(3) & 17.0(4) & -14.4(4)\\
 J1901+0510 & 20210502 & 47.1(2) & 11.9(2) & -11.4(2)  & 72.6(6)  & 22.9(7)  & -21.8(7)  & 36.5(1) & 7.5(1)  & -7.1(1)\\
            & 20210903 & 45.3(3) & 11.9(5) & -11.4(5)  & 68.5(12) & 23.3(15) & -22.2(15) & 33.2(2) & 7.2(2)  & -6.8(2)\\
J1909+0007 & 20210701 & 14.1(1) & 6.4(1)  & 1.9(1)    & 25.1(4)  & 7.9(6)   & -3.0(6)   & 13.5(1) & 6.9(1)  & 2.7(1) \\
J1929+1844  & 20210111 & 16.6(1) & 9.0(1)  & 6.3(1)    & 9.7(3)   & 5.9(4)   & 1.2(4)    & 19.3(1) & 12.8(1) & 8.2(1) \\
            & 20220602 & 13.6(1) & 5.7(2)  & 4.1(2)    & 9.8(2)   & 5.2(3)   & 0.2(3)    & 18.8(1) & 11.6(2) & 7.9(2) \\
			
\noalign{\smallskip}\hline
\end{tabular}
\end{table*}

In order to get a better understanding of the mode-changing phenomenon, several physical interpretations have been proposed, which are generally related to the different states in pulsar magnetosphere. For example, \citet{Wang2007} suggested that redistribution of magnetospheric current flow results in mode changes or nulling. Different geometries might lead to mode changes \citep{Timokhin2010}. The emission state can also be affected by temporal modifications of the local magnetic field structure and strength at the surface of the polar cap \citep{Geppert2021}. Moreover, changes of pulse profiles were found to be closely related to different spin-down rates of the pulsars \citep{Lyne2010}. 

In addition to the change of pulse profiles, the polarization might also be different for emission of different modes. For instance, distinct linear polarization degrees of different emission modes have been detected for pulsars such as PSRs J0738-4042, J0742-2822 and J1938+2213 \citep{Karastergiou2011,Keith2013}. The circular polarization of the two emission modes might also be different, for example, the opposite senses for PSR B0943+10 \citep{Suleymanova1998}. In some cases, the polarization position angles remain consistent between the two emission modes of pulsars such as PSRs J1822-2256, B0329+54, J2321+6024 and J1727-2739 \citep[e.g.][]{Basu2018_J1822m2256, Brinkman2019, Rahaman2021, Rejep2022}. However, the polarization position angle curve of PSR J0614+2229 was found to be shifted along the rotation phase for the two emission modes \citep{Sun2022}, which is probably caused by different emission heights for two modes via the aberration effect \citep{Blaskiewicz1991}. In some cases, such as PSRs J0946+0951, J2006-0807 and J0738-4042, the mode-changing phenomenon is closely related to the orthogonal polarization modes \citep{Suleymanova1998,Rankin2006,Basu2018_J2006m0807,Karastergiou2011}.
\citet{Ilie2020} detected the  polarization difference of the modulated emission observed for the two drifting modes, and suggested that the change of drifting modes is not only caused by in the underlying carousel radius but also related to magnetospheric propagation effects. Therefore, polarization properties are important in probing the physical mechanisms or emission conditions for the mode-changing. 

During the Galactic Plane Pulsar Snapshot (GPPS) survey \citep{Han2021} by the Five-hundred-meter Aperture Spherical radio Telescope \citep[FAST,][]{Nan2006,Nan2011IJMPD}, the sky areas with known pulsars have been observed as a verification of the receiver system, and polarization data are all recorded. Among a large number of pulsars observed, we find four pulsars, PSRs J1838+1523, J1901+0510, J1909+0007 and J1929+1844, exhibiting distinct polarization properties for different emission modes. In this paper, we present the polarization profiles for these mode-changing pulsars. Observation and data processing are briefly presented in Section~\ref{sect:Obse}. In Section~\ref{sect:Resu}, we report on the results for different emission modes of the four pulsars and analyse their polarization properties. Discussion and conclusions are presented in Sections ~\ref{sect:Discu} and ~\ref{sect:Conclu}.

\section{Observations and data reduction}
\label{sect:Obse}

Details of FAST observations of the 4 pulsars are listed in Table~\ref{table:obs}. All data of  but one have been taken by the FAST Galactic Plane Pulsar Snapshot (GPPS) survey \citep{Han2021} using the 19-beam L-band receivers \citep{jth+2020}. The one data set is taken from the open data source in the FAST released achieve for PSR J1901+0510, which is used here mainly for result verification. 
The observations cover the frequency range from 1000 to 1500 MHz, with 4096 or 2048 channels. Full polarization signals, $XX, YY, X^\ast Y$ and $XY^\ast$,  were recorded with a sampling time of 49.152~µs. All pulsars were observed with the zenith angles smaller than 28.5 degrees, so that the FAST has a full gain of $\mathit{G} = 16$~K/Jy. 
The instrument polarization is calibrated with the solutions obtained from On-Off calibration signals around each observation, including the band-pass and polarization response.

\begin{figure*}
\begin{tabular}{cc}
\includegraphics[width=0.41\textwidth]{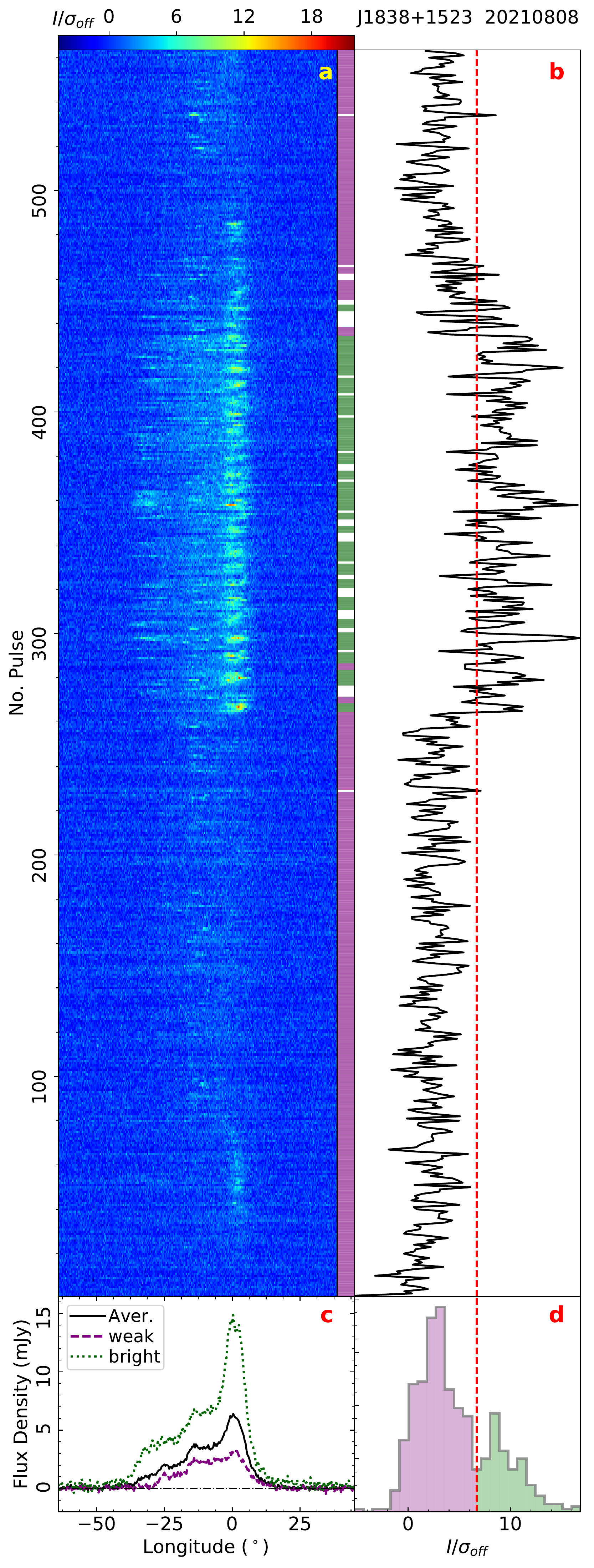}&
\includegraphics[width=0.345\textwidth]{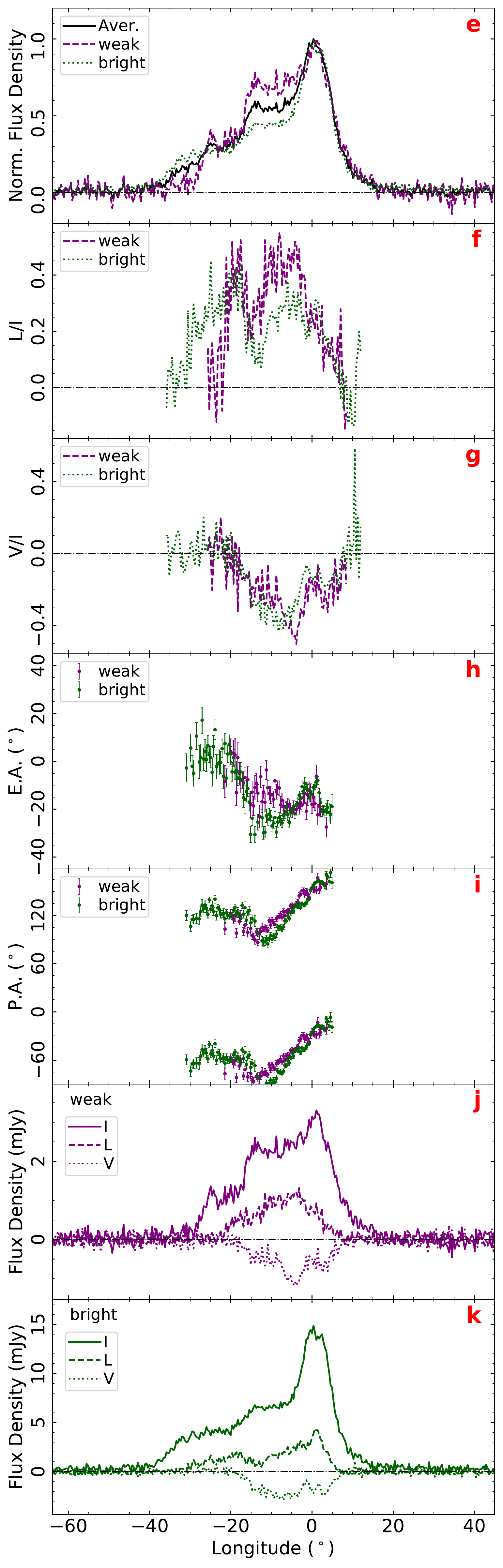}
\end{tabular}
    \caption{
    Individual pulses and the polarization properties for two emission states for PSR J1838+1523, observed by the FAST on 20210808. Individual pulse sequence is shown in {\it the subpanel a}, with the variation of integrated intensity in {\it the subpanel b}. Bright and weak emission modes are labeled by green and purple bars along the sequence, and the pulses that can not be discriminated are left as white. Average profiles of two emission modes as well as all periods are shown in {\it the subpanel c}. The distribution of the integrated pulse intensities is shown in {\it the subpanel d}, which is normalized by $\sigma_{\rm off}$ obtained from the off-pulse window statistics. The two emission modes are discriminated by the threshold indicated by the vertical dashed line. 
    {\it The subpanels e} to {\it k} in the right are comparisons for the total flux density pulse profiles for the weak and bright modes and all averaged periods normalized to their peak {\it (the subpanel e)}, the fractional linear polarization (L/I, {\it the subpanel f}), the fractional circular polarization (V/I, {\it the subpanel g}), the ellipticity angles (E.A.=1/2 arctan(V/L), {\it the subpanel h}), the position angles of linear polarization (P.A., {\it subpanel i}) of the two modes, and also  the averaged polarization profiles (I, L, V), for the weak mode {\it (the subpanel j)} and for bright mode {\it (the subpanel k)}. The PA and EA are plotted only when the total intensity and linear polarization intensity exceed 3$\sigma_{I}$ and 4$\sigma_L$ of the off-pulse region; L/I and V/I are plotted only when the total intensity exceeds  6$\sigma_I$.
    }
    \label{fig:J1838}
\end{figure*}

\begin{figure*}
    \begin{tabular}{cc}
    \includegraphics[width=0.46\textwidth]{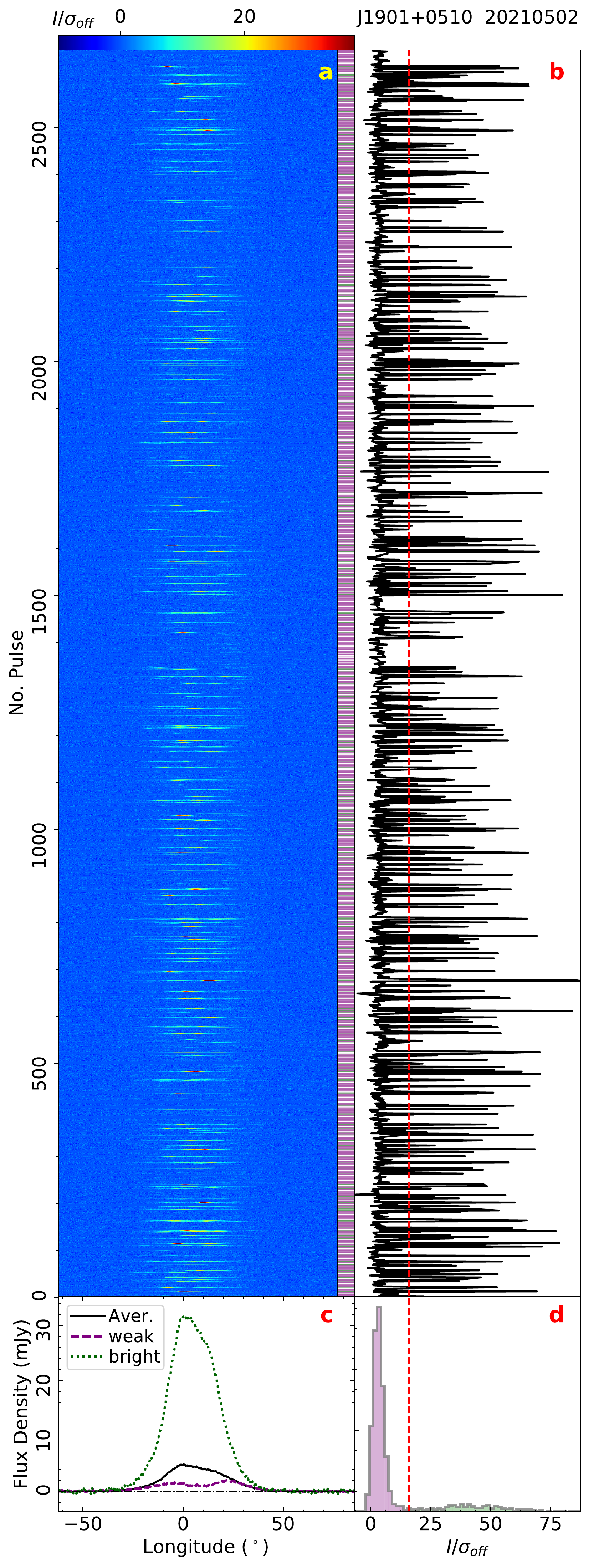}&
    \includegraphics[width=0.388\textwidth]{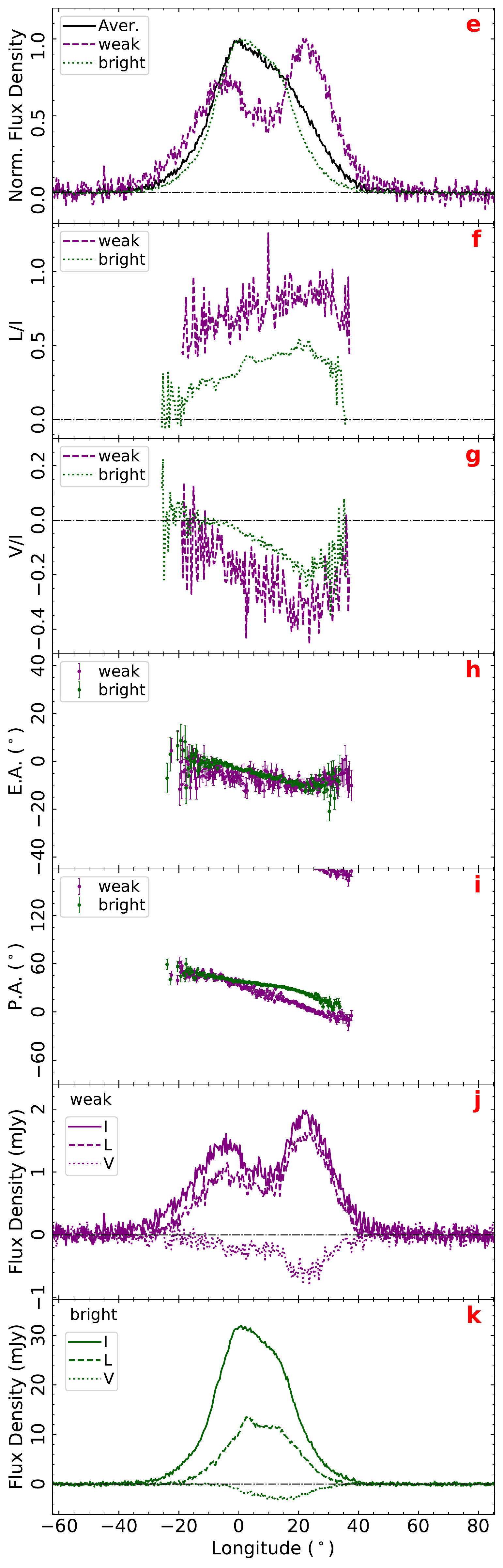}
    \end{tabular}
    \caption{Same as Figure~\ref{fig:J1838} but for PSR J1901+0510 observed on 20210502.
    }
    \label{fig:J1901a}
\end{figure*}

\begin{figure*}
    \begin{tabular}{cc}
        \includegraphics[width=0.46\textwidth]{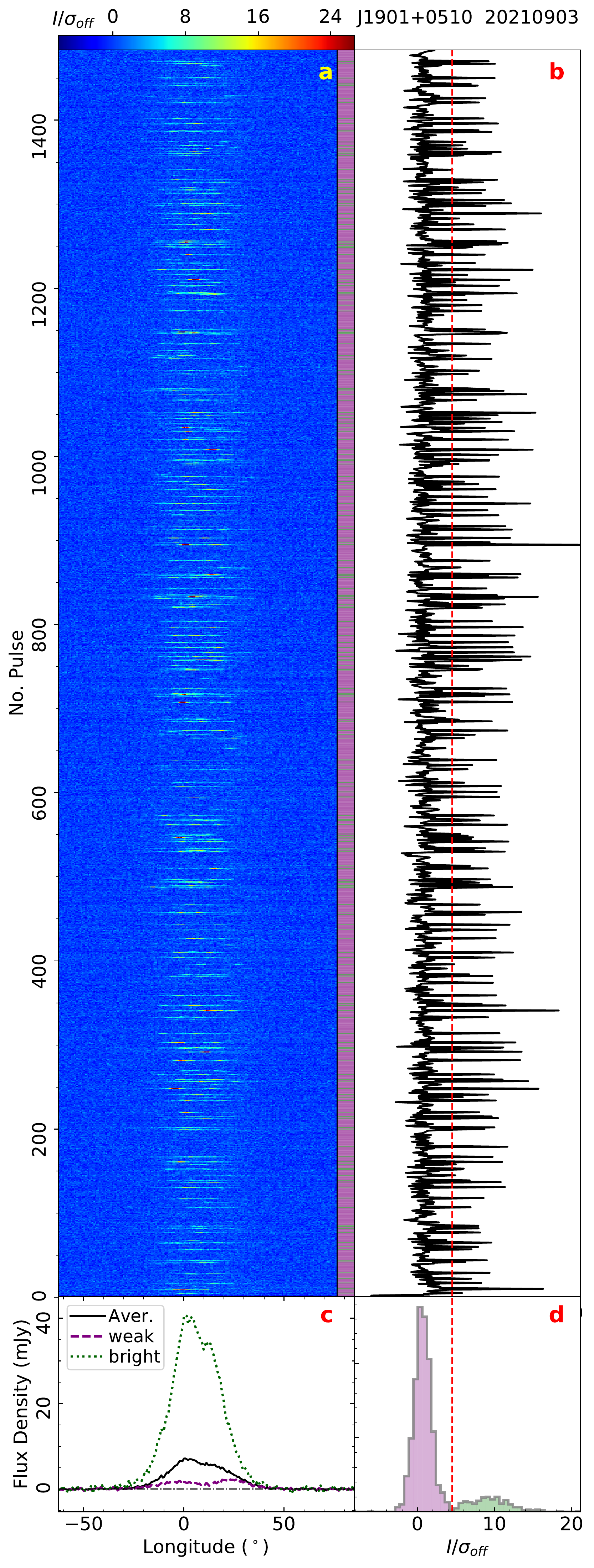}&
        \includegraphics[width=0.388\textwidth]{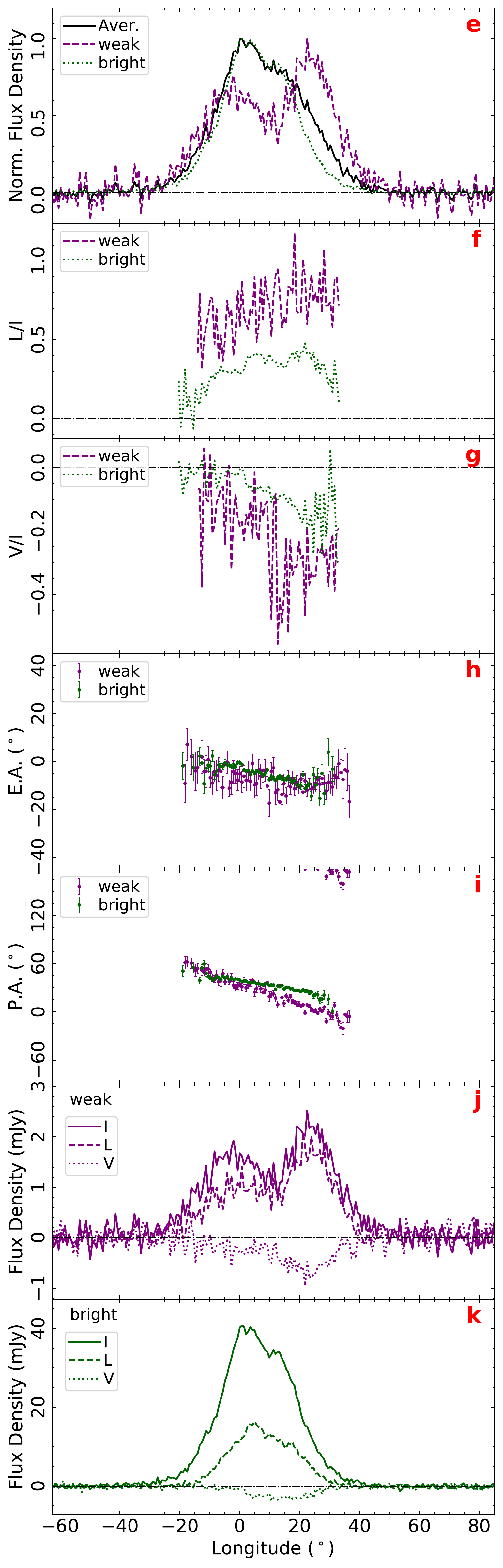}
    \end{tabular}
    \caption{Same as Figure~\ref{fig:J1901a} for PSR J1901+0510 but observed by FAST on 20210903.
    }
    \label{fig:J1901b}
\end{figure*}

\begin{figure*}
    \begin{tabular}{cc}
        \includegraphics[width=0.46\textwidth]{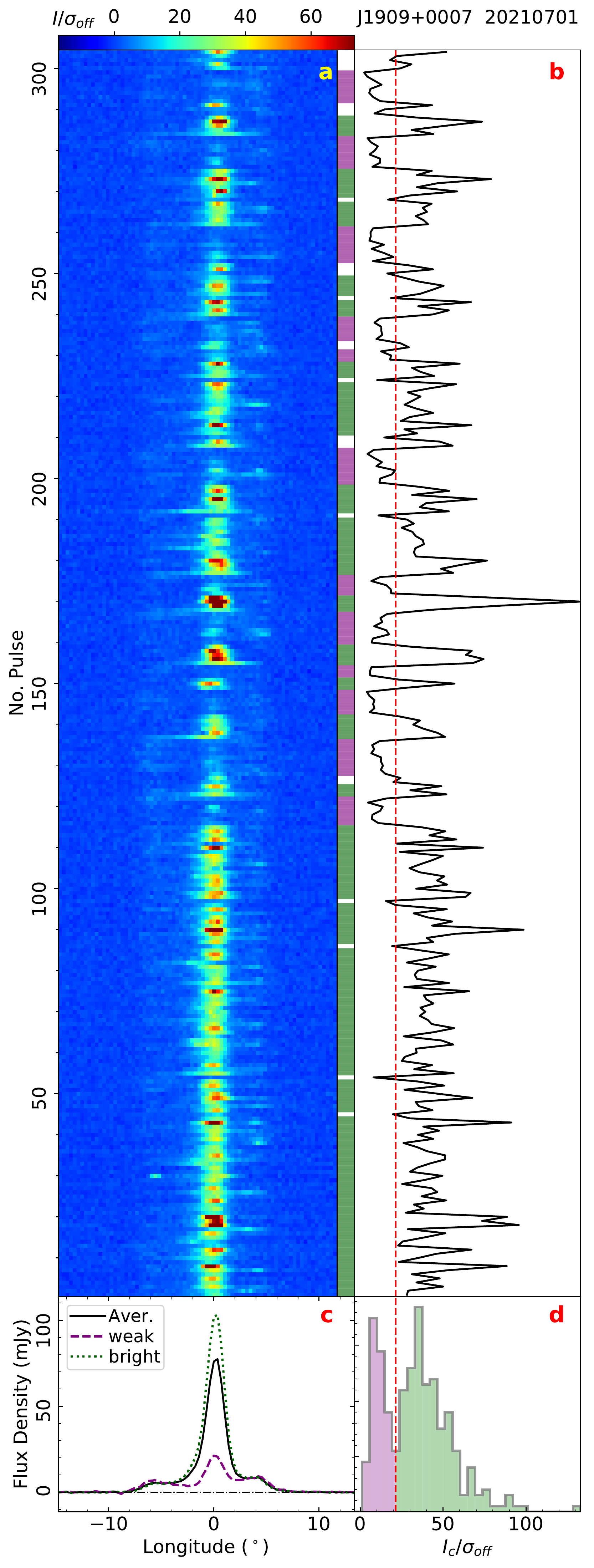}&
        \includegraphics[width=0.388\textwidth]{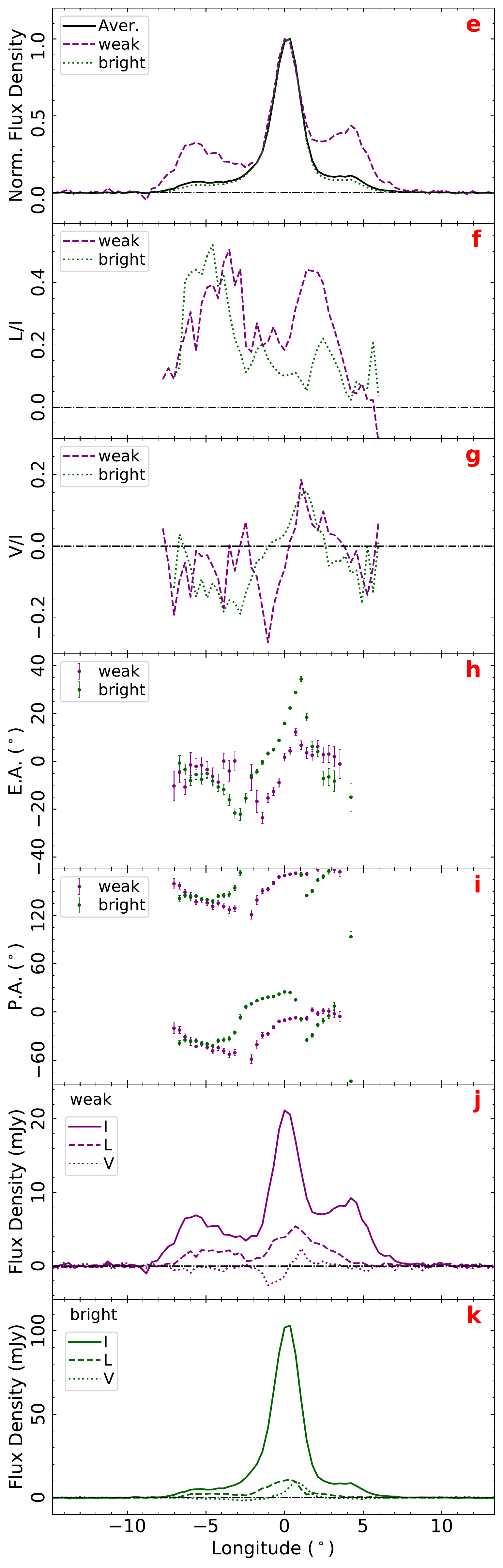}
    \end{tabular}
    \caption{Same as Figure~\ref{fig:J1838} but for PSR J1909+0007 observed by FAST on 20210701.
    }
    \label{fig:J1909}
\end{figure*}

\begin{figure*}
    \begin{tabular}{cc}
        \includegraphics[width=0.46\textwidth]{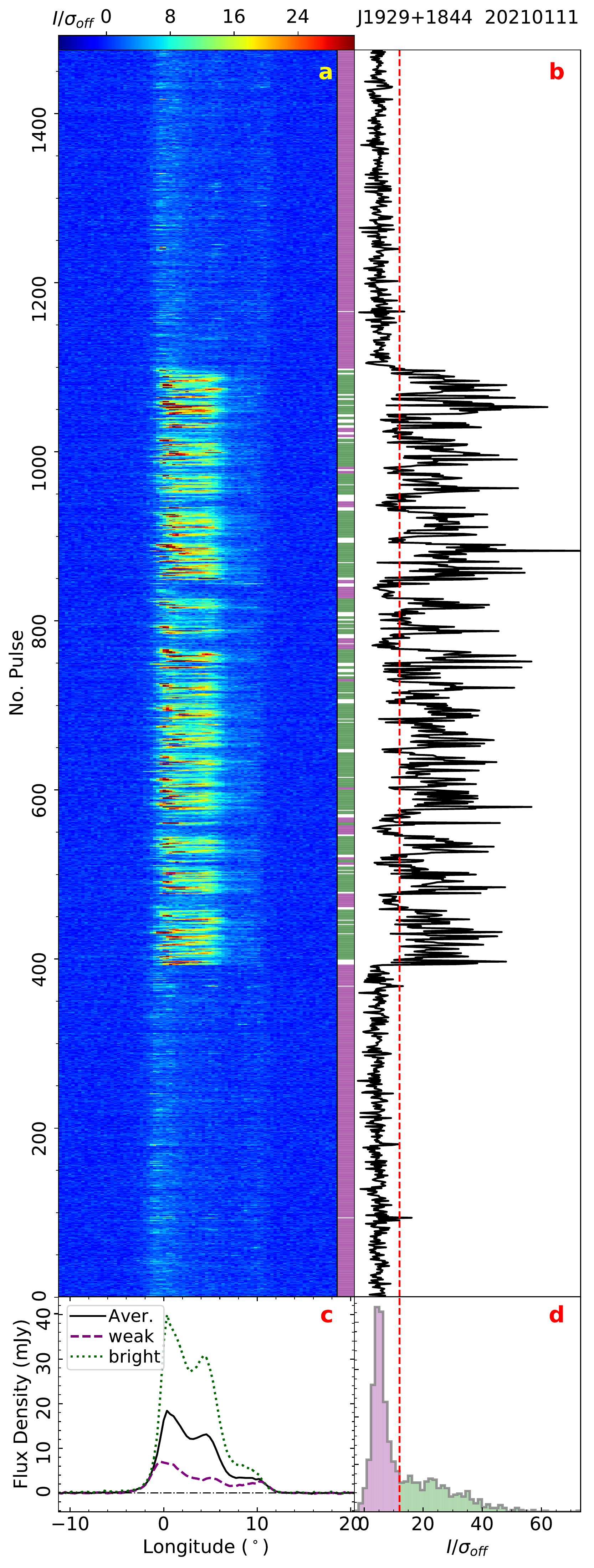}&
        \includegraphics[width=0.393\textwidth]{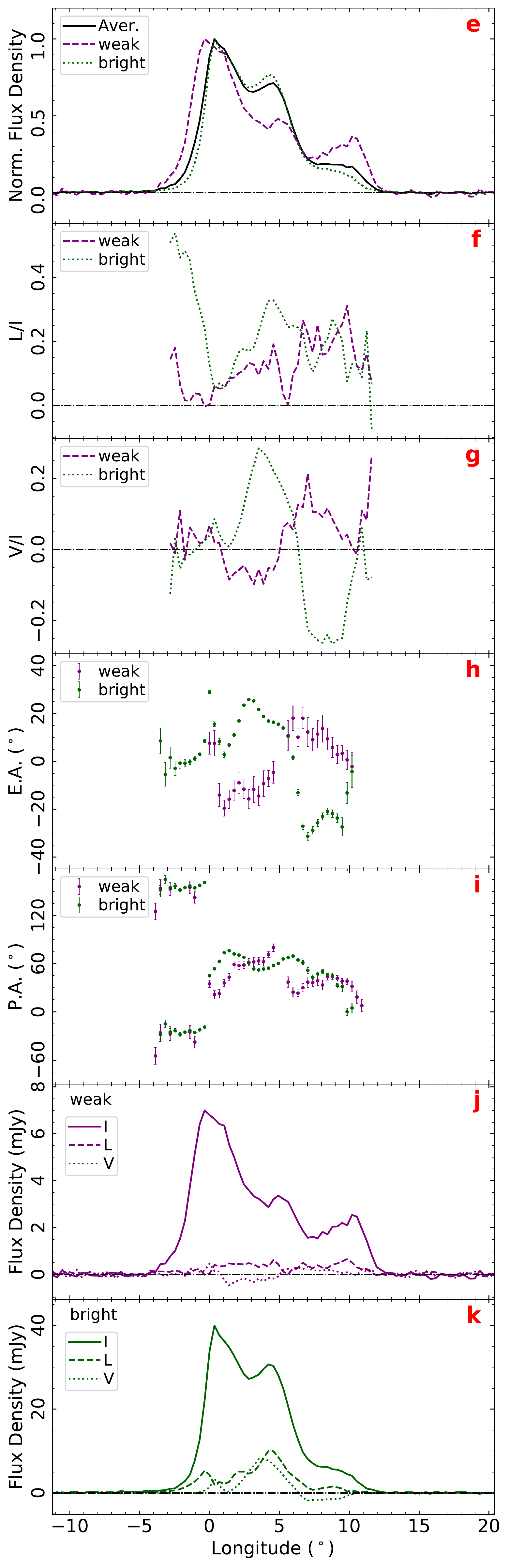}
    \end{tabular}
    \caption{Same as Figure~\ref{fig:J1838} but for PSR J1929+1844 observed on 20210111.
    }
    \label{fig:J1929a}
\end{figure*}

\begin{figure*}
    \begin{tabular}{cc}
        \includegraphics[width=0.46\textwidth]{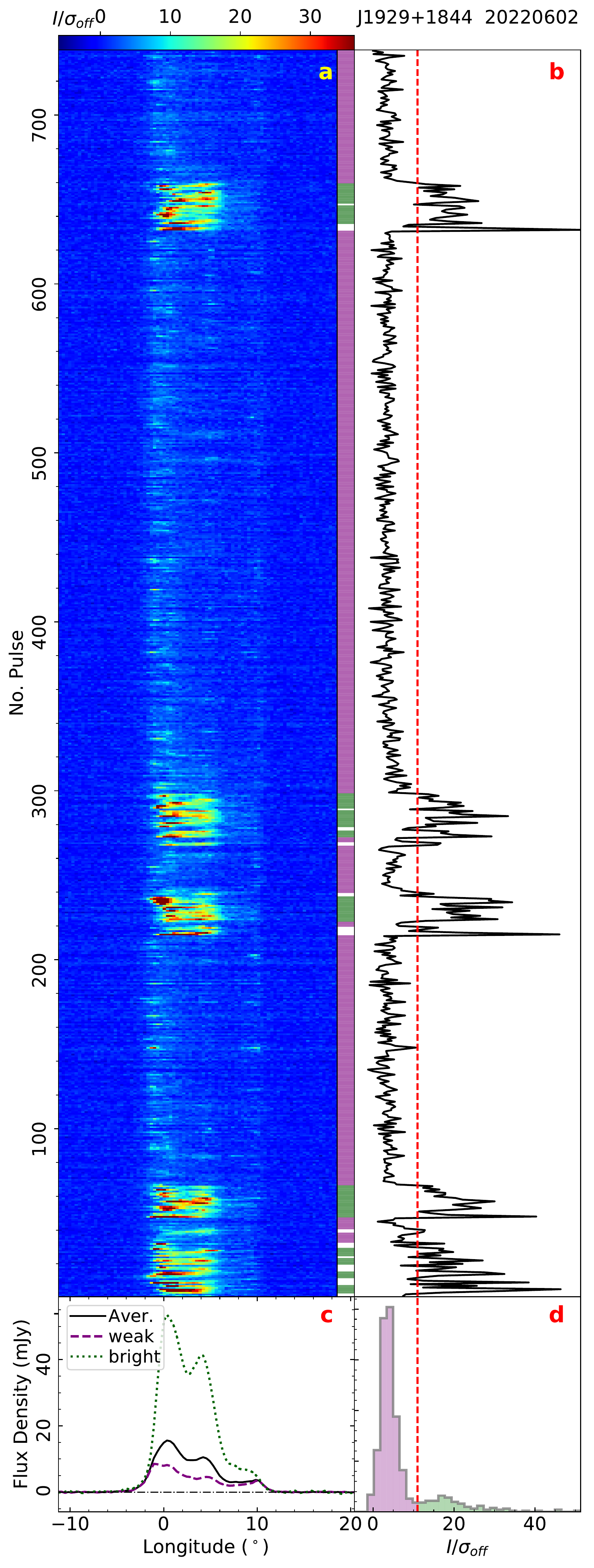}&
        \includegraphics[width=0.393\textwidth]{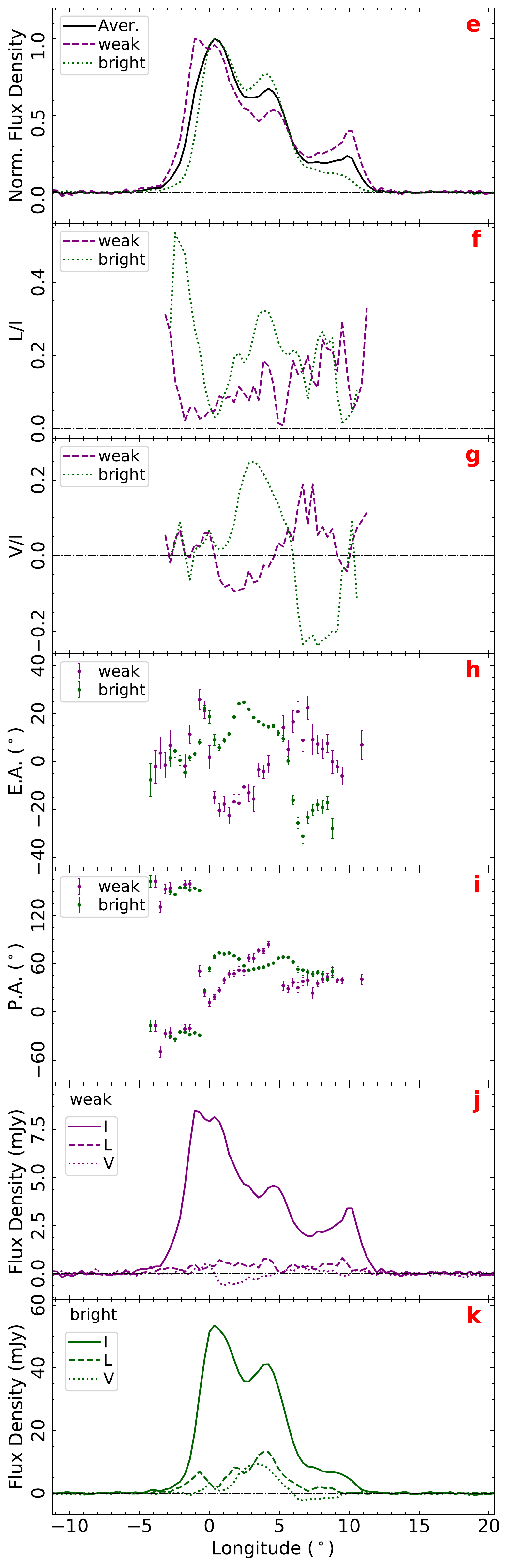}
    \end{tabular}
    \caption{Same as Figure~\ref{fig:J1929a} but observed on 20220602.
    }
    \label{fig:J1929b}
\end{figure*}

Taking ephemerides from the Australia Telescope National Facility Pulsar Catalogue \citep{mht+2005}, we de-disperse and fold data using the open source package DSPSR \citep{van2011}. PSRCHIVE software \citep{Hotan2004} is used to estimate and correct the Faraday Rotation Measure. Single-pulse sequences of all analysed pulsars are shown in Figures~\ref{fig:J1838}-\ref{fig:J1929b}.

\begin{figure*}
\centering

   \begin{subfigure}{0.47\linewidth}
        \centering
        \includegraphics[width=\textwidth]{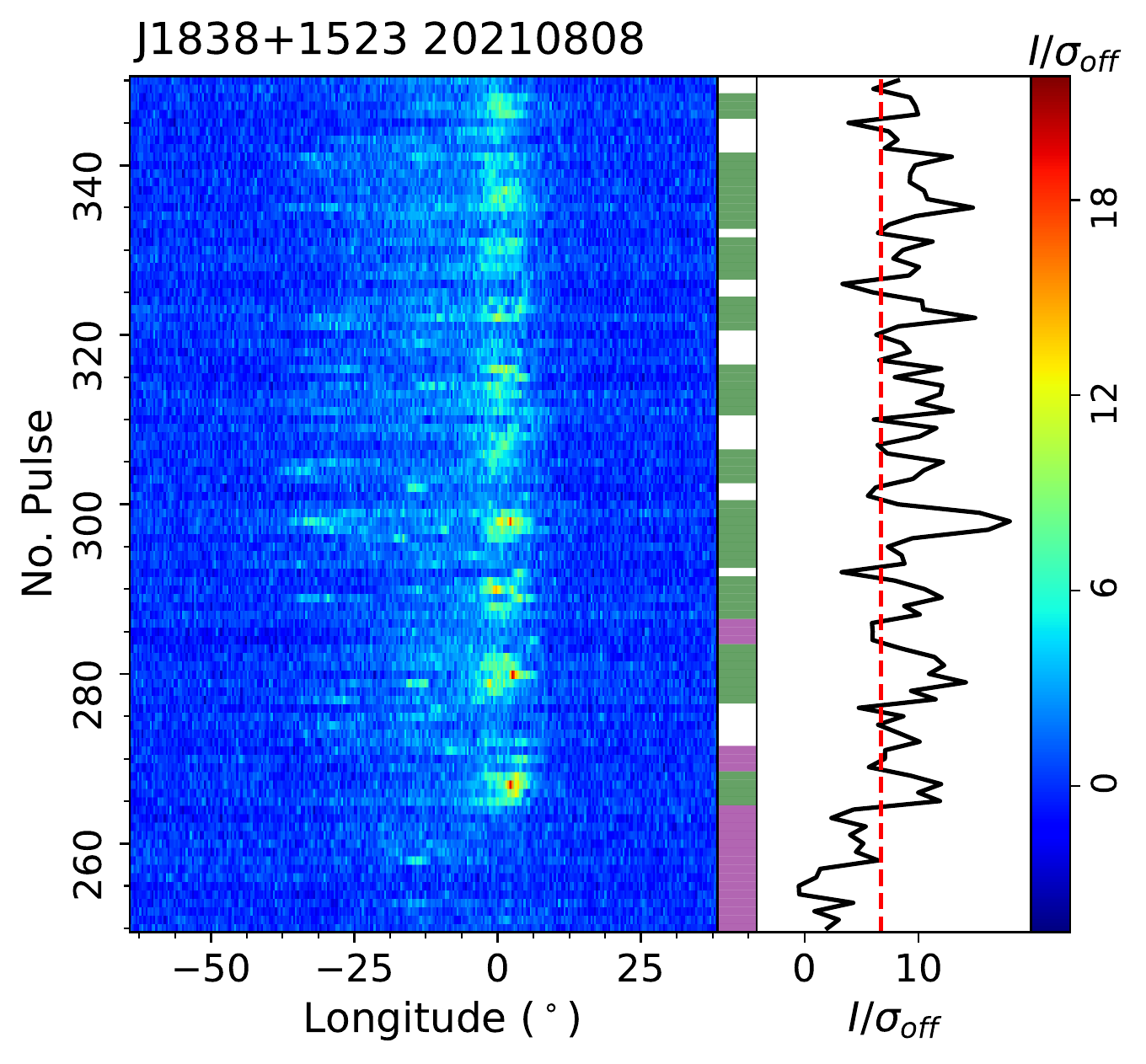}
    \end{subfigure}
    \begin{subfigure}{0.47\linewidth}
        \centering
        \includegraphics[width=\textwidth]{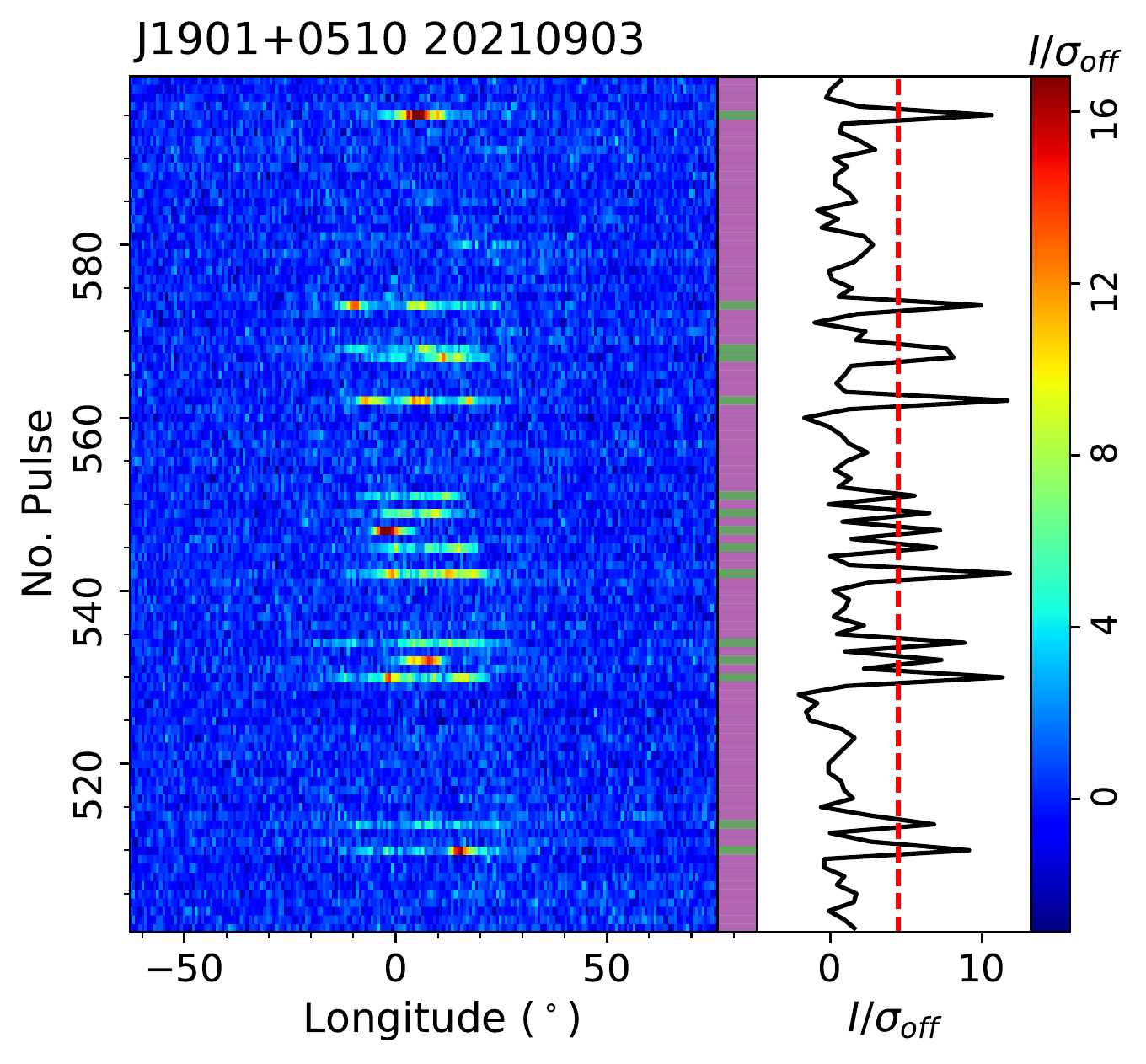}
    \end{subfigure}
    \\
    \begin{subfigure}{0.47\linewidth}
        \centering
        \includegraphics[width=\textwidth]{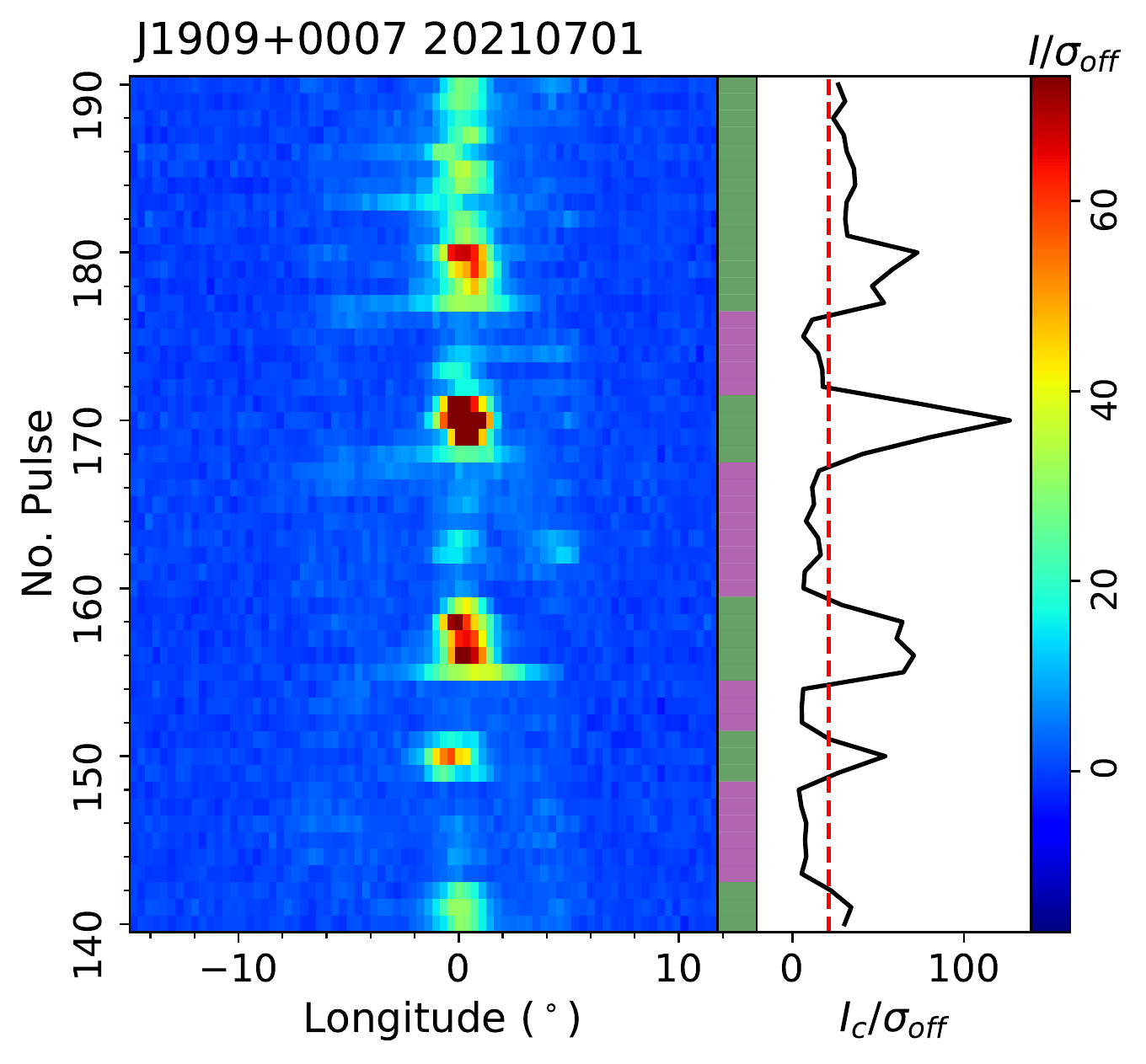}
    \end{subfigure}
    \begin{subfigure}{0.47\linewidth}
        \centering
        \includegraphics[width=\textwidth]{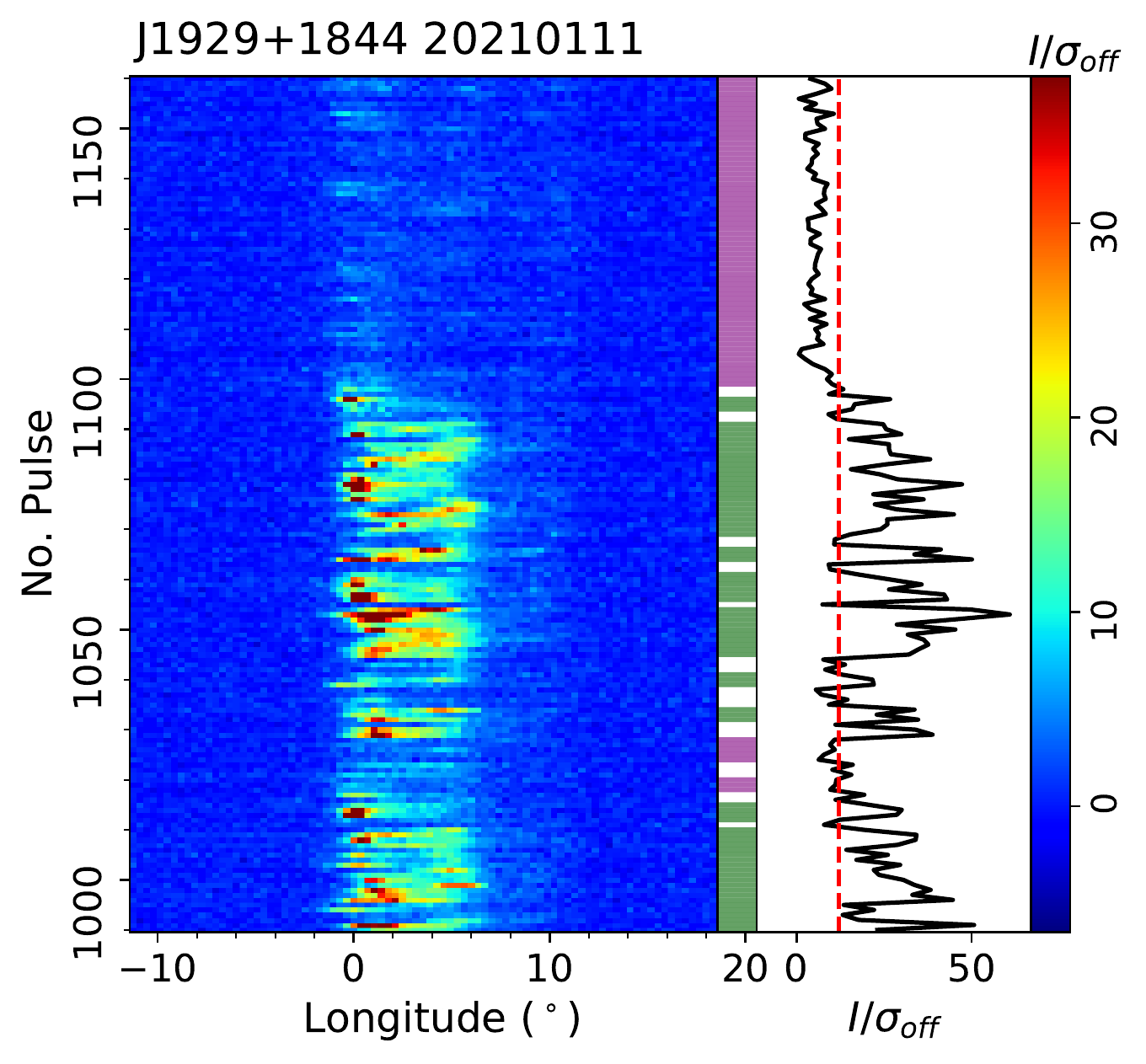}
    \end{subfigure}

\caption{Zoom-in views of some segments of pulse sequences for PSRs J1838+1523, J1901+0510, J1909+0007 and J1929+1844.}
\label{fig:ZoomTPs}
\end{figure*}

\section{Results and analysis}
\label{sect:Resu}

Single pulse sequence was first plotted for each pulsar, as shown in Figure~\ref{fig:J1838} to Figure~\ref{fig:J1929b}. The pulse intensity integrated over the entire pulse window is shown to the right part of the sequence. Two distinct emission states or modes are very apparent in the pulse sequence, and are further verified in the intensity variations. Some zoomed segments  of the pulse sequences and the intensity variations are shown Figure~\ref{fig:ZoomTPs}.

The distribution of integrated pulse intensities shown in Figure~\ref{fig:J1838} to Figure~\ref{fig:J1929b} has a bimodal shape. One can therefore set a threshold to distinguish the bright and weak emission states, as indicated by the red dashed line. Considering the modulation of pulse intensity by random processes, we ignore a mode lasting for only one or two periods  except for the bright pulses of PSR J1901+0510. 

The polarized pulse profiles of two discriminated modes are then obtained by integrating  single pulse data, as shown in the right part of Figure~\ref{fig:J1838} to Figure~\ref{fig:J1929b}. Their polarization properties are also listed in Table~\ref{table:PolFrac}.

\subsection{PSR J1838+1523}

PSR J1838+1523 was found by \citet{Surnis2018} using the GMRT at 325 MHz, and a timing solution was published alongside. 

Here is the first time to report the mode changing of PSR J1838+1523. The FAST GPPS survey observation of this pulsar was made for only 5 minutes on August 08th, 2021 (hereafter 20210808, also for other observation date). Its single pulse sequence exhibits strong emission between pulse number 265 to 460, lasting for about 1.8 minutes, as shown in Figure~\ref{fig:J1838}.  The trailing component is significantly enhanced during the bright state, as shown by the normalized intensities in Figure~\ref{fig:J1838}c. 

Moreover, only the trailing component is strengthened during the pulse period of 40 to 80 in the weak state, which may be another mode, and more better data are needed to make conclusion. 

The linear polarization intensity and the fractional linear polarization vary over the profile components, and they are different in the two states. The polarization position angle (PA) curve in the profile trailing parts in the bright state has a steeper slope than that in the weak state.
Their ellipticity angles show differences in the longitude range from -15$^\circ$ to -6 $^\circ$.
In addition, the strongest trailing component in the bright state seems to have flux modulations (see Fig.~\ref{fig:ZoomTPs}), indicating of plasma activity in the pulsar magnetosphere.

\subsection{PSR J1901+0510}

PSR J1901+0510 was discovered by the Parkes multibeam pulsar survey \citep{Hobbs2004}. 

We got a FAST observation for 15 minutes of PSR J1901+0510 on 20210903 during the confirmation observation of a pulsar candidate through one of the 19 beams. As shown in Figures~\ref{fig:J1901b}, the pulse sequence shows that its bright and weak states switch very frequently (see the zoomed-in single pulse sequence in Figure~\ref{fig:ZoomTPs}). The bright state is maintained for only one or a few periods, while the radio emission is very weak for most periods. 

To verify such a quick mode-switching, we take an archived FAST data observed on 20210502 for 27 minutes, and the results are shown in Figure~\ref{fig:J1901a}. The switching between both emission states is then confirmed, and the polarization results are consistent in the two observations. The intensity distributions exhibit two distinct groups in both Figures~\ref{fig:J1901a} and \ref{fig:J1901b}, which correspond to the bright and the weak states. 

Pulsar polarization profiles for the two states are quite different. As shown in Figures~\ref{fig:J1901a} and \ref{fig:J1901b}. The pulse profiles on the weak state exhibit two distinct components, but the component separation of the bright state is too small to have two peaks. In addition, the polarization position angle curves of the weak state on the two observation sessions have a steeper slope compared with those on the strong state, with a much larger deviation in the trailing component. The fractional linear and absolute circular polarization in the weak state are systematically higher than those for the bright state.

\subsection{PSR J1909+0007}

PSR J1909+0007 (B1907+00) was discovered in the low latitude pulsar survey by the Mark 1A radio telescope at Jodrell Bank \citep{Davies1973}.

Here is the first time to report the mode changing of this pulsar based on a 5 minutes FAST GPPS survey observation conducted on 20210701. Its leading and trailing components are occasionally brightened, whereas the central component is occasionally much weakened, as shown in Figure~\ref{fig:J1909}a. The central component exhibits two distinct distributions of intensities (see Figure~\ref{fig:J1909}d), which enables the classification of single pulses into bright and weak states.

Pulsar profiles in the two states of the central component have different polarization features, as shown in the right part of Figure~\ref{fig:J1909}. Intensity and polarization of the leading and trailing components remain almost unchanged during the switching of the two states, but the central component switches to the very bright state (see Figure~\ref{fig:J1909}c). Effectively, the two shoulder components are enhanced in the weak mode if these profiles are scaled by the peak flux density. The polarization position angles are different by about 45$^\circ$, and the fractional linear polarization is also reduced in the bright mode, both of which induce the difference of elliptic angles. These variations might be caused by the competition of the orthogonal polarization modes.

\subsection{PSR J1929+1844}

PSR J1929+1844 (B1926+18) was found by Arecibo at 430 MHz \citep{Hulse1975}. It is a canonical example of a mode-changing pulsar. The mode changing phenomenon was previously reported by \citet{Ferguson1981}. In its "abnormal mode", all three components become bright and intensity of the central component is extremely enhanced \citep{Ferguson1981,Weisberg1986,Nowakowski1994}. But polarization behaviours of these modes were not reported before.  

We got two FAST observations for 30 minutes and 15 minutes on 20210111 and 20220602, respectively, during the FAST GPPS survey verification observations for a pulsar candidate. The bright and weak states of this pulsar are clearly shown in the single pulse sequences in Figures~\ref{fig:J1929a} and \ref{fig:J1929b}. A zoomed in view of single pulse sequence is displayed in Figure~\ref{fig:ZoomTPs} for pulses between 990 and 1160 in Figure~\ref{fig:J1929a}. The bright state can last for several tens to several hundreds of periods. During which, the pulsar might occasionally switches back to its normal mode. The bright and weak pulses are interspersed especially in the bright state, which leads the intensity histogram of both states to be overlapped in Figures~\ref{fig:J1929a}d and \ref{fig:J1929b}d. 

As shown in Figures~\ref{fig:J1929a} and \ref{fig:J1929b}, the weak mode shows a wider pulse profiles, and the pulse window extends from a longitude of about -3.8$^\circ$ to 12.5$^\circ$. There is a notch at the longitude of 0$^\circ$ for the weak state. When the pulsar switches to the bright state, only the pulses within the phase range of -1.0$^\circ$ to 10.5$^\circ$ are enhanced. It leads the notch exhibiting in the weak state to be filled. Moreover, different components are enhanced differently, with the central component growing the most and the trailing component the least. This behaviour is consistent with the previous observations at 430~MHz, although the central component is not dominating over the leading one at this FAST observation frequency.

The polarization behaviours of two emission states are remarkable. Around the pulse peak phase nearly at the longitude of $0^\circ$, the orthogonal polarization modes with a $90^\circ$ jump is seen in polarization position angle curves for both states. For emission prior to the peak phase, the polarization position angles remain almost consistent for the states, though the fractional linear polarization of the bright state is much higher than that of the weak one. For the emission after the peak phase, the position angle curves are very different for the two states, with a possible orthogonal polarization mode jump at the longitude of 5.0$^\circ$ in the weak mode. The most striking is the completely reversed senses of circular polarization for the two states, from the left hand to the right hand for the bright state, to just the opposite for the weak state.
The degrees of linear and absolute circular polarization  are low for the whole pulse profile in the weak mode compared with that of the bright one (see Table~\ref{table:PolFrac}).
The variation of ellipticity angles reflect both the reversals of circular polarization senses and the very different PA curves for the two states.

\begin{figure}
\centering
\setlength\tabcolsep{0pt}
\begin{tabular}{cc}
\includegraphics[width=0.39\textwidth]{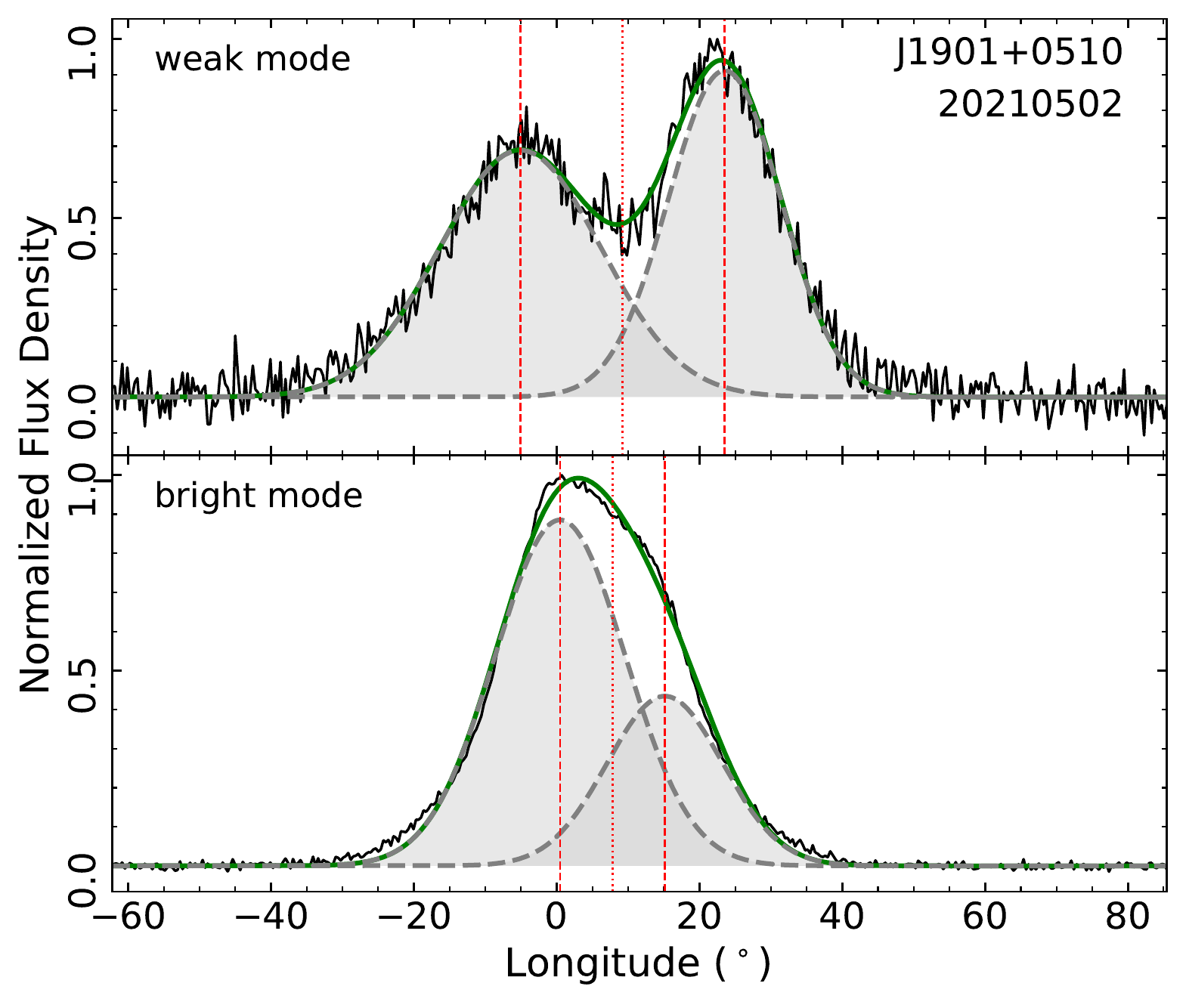}\\
\includegraphics[width=0.39\textwidth]{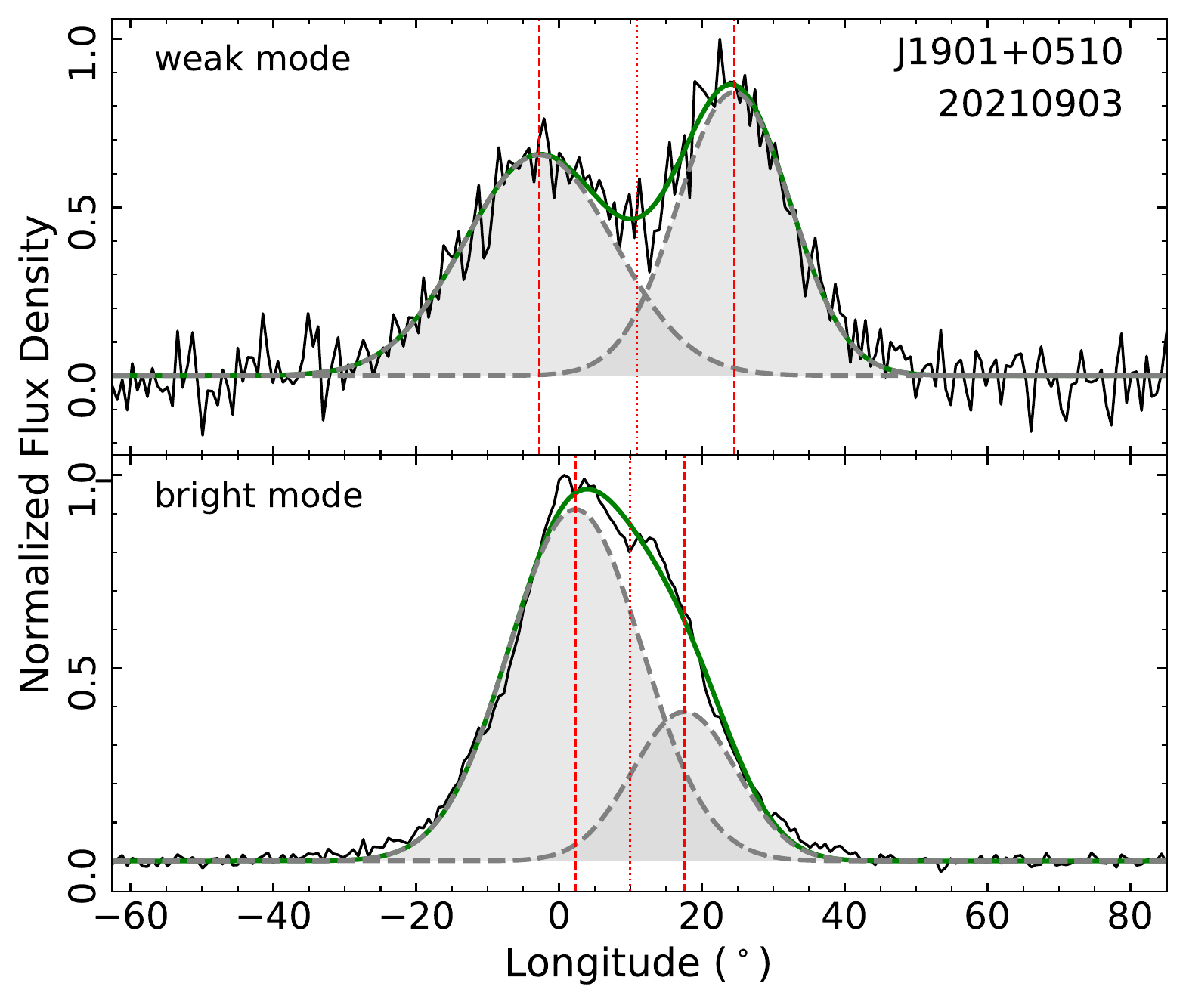}
\end{tabular}
\caption{Component separation of the pulse profiles of J1901+0510 for its two emission states observed on 20210502 and 20210903. The black and green solid lines are integrated pulse profiles and their fittings with two-Gaussian function. Fitted Gaussian curves corresponding to components are plotted with the dashed grey lines. The red vertical dashed lines indicate the peak positions of the two components, and the central dotted vertical line represents the profile center. The two components are separated by about 28$^\circ$ for the weak state, but only about 15$^\circ$ for the bright state. }
\label{fig:J1901+0510_comp}
\end{figure}

\begin{figure}
    \begin{subfigure}{\linewidth}
        \centering
        \includegraphics[width=0.84\textwidth]{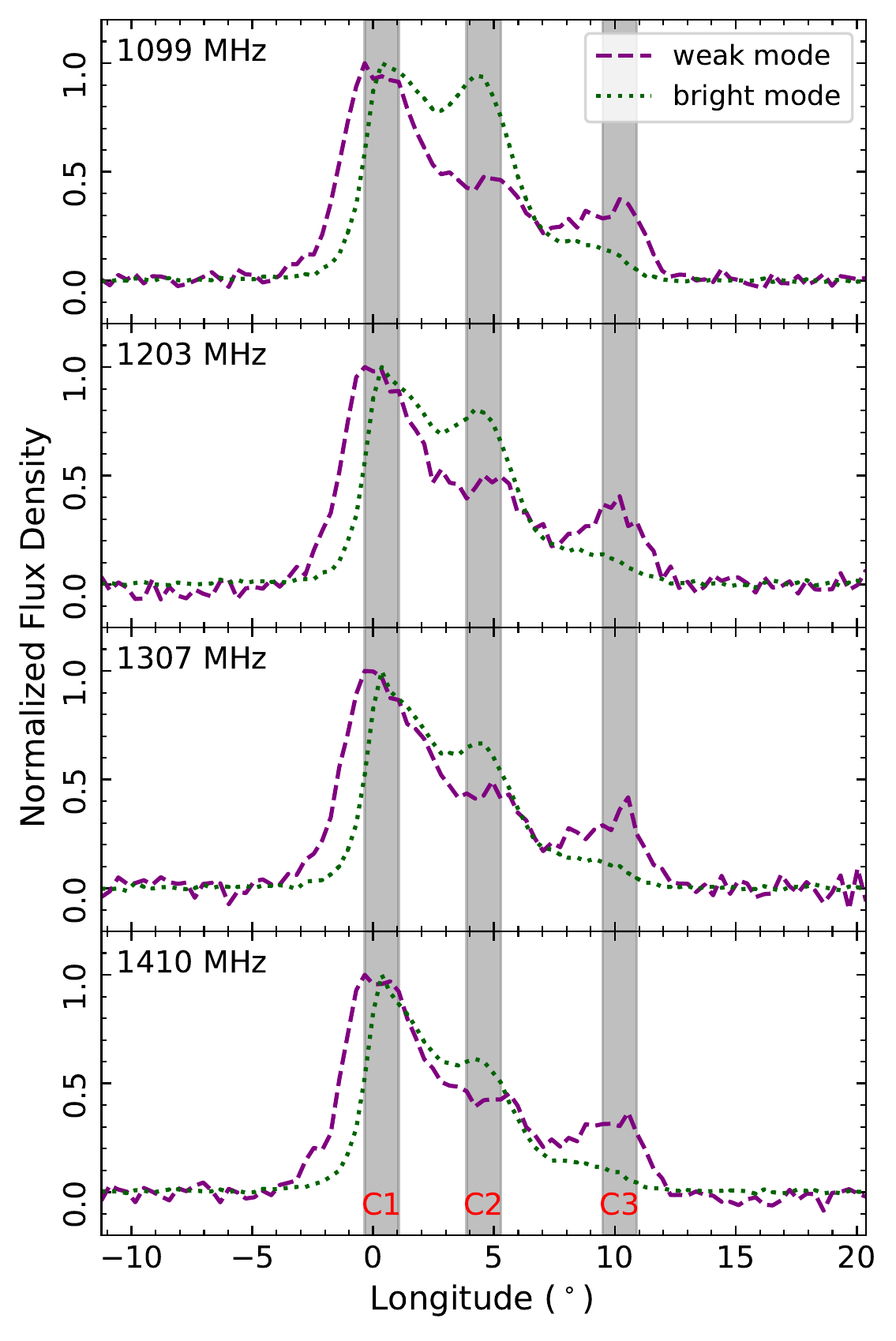}
        \caption{Total intensity profiles of J1929+1844 in 4 sub-bands in two modes}
    \end{subfigure}
    \begin{subfigure}{\linewidth}
        \centering
        \includegraphics[width=0.84\textwidth]{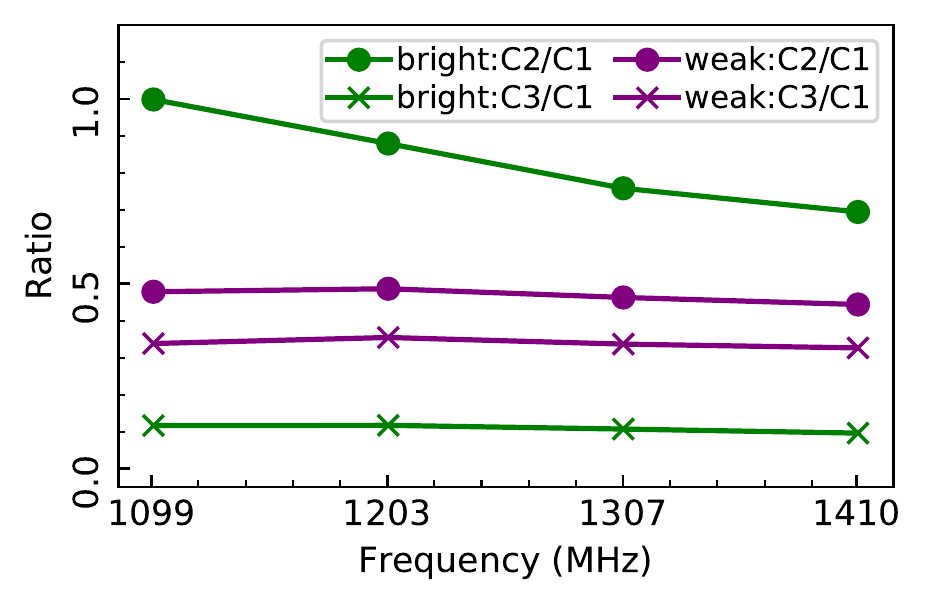}
        \caption{Intensity variations of the main components}
    \end{subfigure}
    \caption{The total intensity profiles of PSR J1929+1844 in the 4 subbands (1099, 1203, 1307 and 1410 MHz) of FAST observations on 20210111 in the two states, and the peak ratios of main components  (C1, C2 and C3) compared to the peak value of the leading peak (C1).}
    \label{fig:J1929_freq}
\end{figure}

\section{Discussion}
\label{sect:Discu} 

As presented above, the pulse polarization profiles, either the PA curves, or the fractional linear polarization, or the fractional circular polarization or even the senses, can be very different for the two emission states of PSRs J1838+1523, J1901+0510, J1909+0007 and J1929+1844, which might be related to physical processes or conditions within pulsar magnetosphere.

\subsection{Different Components in two emission states}

Profile components of PSR J1901+0510 vary the most during its state switching. As shown in Figure~\ref{fig:J1901+0510_comp}, the component separation is about 15$^\circ$ for the bright state but widened to 28$^\circ$ for the weak state. Moreover, the profile center of the weak state is delayed by approximately 1.2$^\circ$ compared to that of the bright state.  

Pulsar radio emission is generally believed to be generated from the open magnetic field lines of a dipole magnetosphere. Assuming that the emission of two states are produced by the same bundle of relativistic particles, its wide profile together with the wide component separation suggests that the emission of weak state originates from a larger height of pulsar magnetosphere compared to the bright state. Moreover, according to the aberration and retardation effects, the profile originating from a higher altitude is shifted towards an earlier phase compared with that emission from a lower altitude \citep{Gangadhara2001}. However, the profile of the weak state is delayed compared with that of the bright one. The contradiction might be caused by the asymmetry of the profile components within the pulsar conal emission beam \citep{xu97}. Otherwise, emission of two states should originate from different bundles of relativistic particles, and the emission on weak state is generated from a lower altitude. 

In addition to the time-dependent features for PSR J1929+1844, the bright and weak states of the pulsar also have different frequency evolution. Figure~\ref{fig:J1929_freq} shows its profiles of the two states across our FAST observing frequency band from 1000 to 1500~MHz. The profiles remains almost unchanged with frequency for the weak state, while significant changes are observed for the bright state, mainly because the central component becomes stronger at lower frequencies and finally dominates at 430~MHz \citep{Hulse1975}. 

\subsection{Different linear polarization in two emission states}

As demonstrated in \citet{wh16}, the fractional linear polarization can be of various degrees, depending on the interaction of the orthogonal polarization modes (X and O mode). If both modes have comparable intensities, the net fractional linear polarization will approach zero. The fraction will be high, if one mode dominates over the other as caused by rotation induced mode separation or refraction \citep{wh16}.

The fractional linear polarization very differs in the two emission states of PSR J1901+0510. As listed in Table~\ref{table:PolFrac}, it reaches about 70\% for emission in the weak state.
As analyzed in the previous section, emission of the weak state of PSR J1901+0501 is likely to be originated from a smaller emission height. Where the refraction might play an important part in separating the modes, and hence the fractional linear polarization could be high. 

The strong and weak states of PSR J1909+0007 might be dominated by different polarization modes for the central component. Competition of polarization modes may be different in the two emission states, causing distinct fractional linear polarization.
The main polarization mode of PSR J1929+1844 at the phase prior to the profile peak becomes more dominated when the pulsar switches to the bright state, causing a larger fractional linear polarization.

\subsection{Different polarization position angle curves of two emission states}

The position angles of linear polarization exhibit different variations in the two emission states for these 4 pulsars. One type is the change in the steepness of PA curves for PSRs J1838+1523 and J1901+0510, and the others include the shift of PA curves for PSR J1929+1844 and very different curves for PSR J1909+0007. These different PA curves in two modes may have different physical origins. 

The change of steepness of PA curves might be caused by the aberration effect, depending on emission heights. If emission of two states originates from different heights, their PA curves at the same phase have different gradients, as demonstrated in \citet{Blaskiewicz1991} and \citet{wwh12}. For example, the PA curve of PSR J1901+0510 in its weak state has a larger gradient. The slightly delayed phase of the profile center indicates a smaller emission height for the weak mode.

The complicated shifts of PAs might be resulted from the superposition of emission from different parts of pulsar magnetosphere. The emission from the same region will have polarization vectors either perpendicular to or within the local magnetic field line plane. If emission originates from different parts in the magnetosphere, the magnetic field line planes are not parallel. Superposition of them will result in the variation of PAs. The central components of the bright state of PSRs J1909+0007 and J1929+1844 are likely produced in another parts of pulsar magnetosphere compared to the weak state, as demonstrated by its distinct spectrum in the previous section, and the superposition of them leads to the variation of PAs.

\subsection{Different circular polarization for two emission states}

The different fractional circular polarization of PSR J1901+0510 and the sense reversal of circular polarization around the central component of PSR J1929+1844 in the two emission states might be caused by different emission processes or propagation effects \citep[e.g.][]{wlh10,wwh12}. Variations of density gradients of the relativistic particles will lead the fractional circular polarization to be changed, depending on the sight line \citep{wwh12}.

\section{Conclusions}
\label{sect:Conclu}

In this paper, we investigate on the polarization features of four pulsars that have weak and bright states. For PSRs J1838+1523, J1901+0501 and J1909+0007, this is the first time to report the mode-changing features. For PSR J1929+1844, the polarization properties of both states are first investigated in this paper.
For these two emission states, significant differences have been observed in the profile morphology, the position angles of linear polarization, the fractions of linear and circular polarization and its senses. For PSR J1838+1523, the profile morphology, the steepness of its position angle curve and its fractional linear polarization all change between two emission modes. For PSR J1901+0510, the profile morphology, the steepness of position angle curves, and the fractional linear and circular polarization are changed for the whole pulse profile. For PSR J1909+0007, the central component exhibits very different intensity, position angle, and fractional linear polarization. For PSR J1929+1844, the central component of its bright state has different frequency evolution, and exhibit significant variation of position angles, fractional linear polarization and the sense reversals of circular polarization. 

These different properties of the two emission modes give insights into the physical processes and conditions of pulsar magnetosphere. Rearrangement of the relativistic particles probably plays an important part in causing the state switchings. Its influences, aberration effect, density gradient of relativistic particles together with the superposition of orthogonal polarization modes may result in various observational features during emission state switchings. 

\section*{Acknowledgements}
P. F. Wang is supported by the National Key
R\&D Program of China (No. 2021YFA1600401 and
2021YFA1600400), National Natural Science Foundation
of China (No. 11873058, 12133004) and the National
SKA program of China (No. 2020SKA0120200). 
J.~L. Han is supported by the National Natural Science Foundation of China (NSFC, Nos. 11988101 and 11833009) and the Key Research
Program of the Chinese Academy of Sciences (Grant No. QYZDJ-SSW-SLH021).

\section*{Data Availability}

Original FAST observational data are open sources in the FAST Data Center according to the FAST data 1-year protection policy. The folded and calibrated data in this paper can be obtained from authors by kind requests.



\bibliographystyle{mnras}
\bibliography{DiffPolModingPSRs.bib} 








\bsp	
\label{lastpage}
\end{document}